\documentclass[conference]{IEEEtran}
\usepackage[utf8]{inputenc}
\usepackage[cmex10]{amsmath}
\usepackage{upgreek}
\usepackage{booktabs}
\usepackage{epsfig}
\usepackage{latexsym}
\usepackage{multirow}
\usepackage{stfloats}
\usepackage{epstopdf}
\usepackage{color}  
\usepackage{tabularx} 
\usepackage{algorithm}
\usepackage{amssymb}
\usepackage{enumerate}
\usepackage{array}
\graphicspath{{./Figures/}}
\usepackage{color}
\usepackage{bbm}
\usepackage{bm}
\usepackage{cite}
\usepackage[tight,footnotesize]{subfigure}
\usepackage{balance}
\usepackage{mathrsfs}
\usepackage{verbatim}
\usepackage{dsfont}
\usepackage{verbatim}
\usepackage{tikz}
\usepackage{setspace}
\usepackage{diagbox}
\usepackage{multicol}
\usepackage{environ}
\usepackage{tikz}
\usepackage{amsmath}
\usepackage{stfloats}
\usepackage{algorithm}
\usepackage{algpseudocode}
\usepackage{amsmath}
\usepackage{graphics}
\usepackage{epsfig}
\allowdisplaybreaks[4]
\usepackage{amsmath}
\usepackage{amsthm}
\usepackage{authblk}

\def\BibTeX{{\rm B\kern-.05em{\sc i\kern-.025em b}\kern-.08em
		T\kern-.1667em\lower.7ex\hbox{E}\kern-.125emX}}
\begin{document}
	\title{Molecular Absorption Effect: A Double-edged Sword of Terahertz Communications
	%: Challenges and Opportunities %Open Problems, and Design Guidelines
	}
	\author{Chong Han, Weijun Gao, Nan Yang, and Josep M. Jornet}
	\maketitle
	\thispagestyle{empty}
	\begin{abstract}
	Communications in the terahertz band (THz) (0.1--10~THz) have been regarded as a promising technology for future 6G and beyond wireless systems, to overcome the challenges of evergrowing wireless data traffic and crowded spectrum. As the frequency increases from the microwave band to the THz band, new spectrum features pose unprecedented challenges to wireless communication system design. The molecular absorption effect is one of the new THz spectrum properties, which enlarges the path loss and noise at specific frequencies. This brings in a double-edged sword for THz wireless communication systems. 
	On one hand, from the data rate viewpoint, molecular absorption is detrimental, since it mitigates the received signal power and degrades the channel capacity. On the other hand, it is worth noticing that for wireless security and covertness, the molecular absorption effect can be utilized to safeguard THz communications among users.  
	In this paper, the features of the molecular absorption effect and their impact on the THz system design are analyzed under various scenarios, with the ultimate goal of providing guidelines to how better exploit this unique THz phenomenon. Specifically, since the molecular absorption greatly depends on the propagation medium, different communication scenarios consisting of various media are discussed, including terrestrial, air and space, sea surface and nano-scale communications. Furthermore, two novel molecular absorption enlightened secure and covert communication schemes are presented, where the molecular absorption effect is utilized as the key and unique feature to boost security and covertness. 
	\end{abstract}
	
	\section{Introduction}
    The rapid increase in worldwide wireless traffic demands new advancements in wireless communications and networking technologies. To break the bottleneck of the limited bandwidth resource, a new spectral band is awaited to be explored to achieve ultra-high-speed wireless transmissions in future 6G and beyond systems. Terahertz (THz) communications, with frequency ranging from 0.1~THz to 10~THz, is envisioned to support low-latency and Terabits-per-second wireless links. 
    Moreover, the potential of THz communications for enhancing information security has attracted great attention due to its high directivity and high path loss~\cite{akyildiz2014terahertz,ma2018security,myarticle1,myarticle2}.
 
    \textit{Molecular absorption} is one of the key THz spectrum features, as modeled by the International Telecommunication Union (ITU)~\cite{ITUR}, with the following definition. When a THz electromagnetic (EM) wave propagates in a non-vacuum medium, it induces resonances within specific molecules along the propagation path, and the resulting energy loss in the EM wave is referred to as the molecular absorption effect. From the communications perspective, the molecular absorption effect brings about additional attenuation and noise, which are known as the  \textit{molecular absorption loss} and \textit{molecular absorption noise}, respectively~\cite{jornet2011channel}. The molecular absorption effect creates path loss peaks in the THz spectrum, which imposes a significant, and in fact mostly detrimental, effect on THz system design. {Since this effect is significantly weak in the micro-wave and millimeter-wave bands, it is one of the most unique properties in the THz band.} 
    
    A widely adopted strategy to counteract the molecular absorption effect of the THz spectrum is to avoid transmitting signals under path loss peaks~\cite{akyildiz2014terahertz}. Between such peaks, multiple distance-dependent spectral windows emerge in the THz band, where the molecular absorption is negligibly weak. 
		{Signal transmitted in these spectral windows experiences the minimum molecular absorption loss, which is beneficial for providing reliable wireless transmission.}
    Thus, THz transmission is recommended in spectral windows only. 
    Combined with the distance- and frequency-selectivity, the high environmental dependency of the molecular absorption effect poses a new challenge to the regulation of spectral windows in various communication scenarios and geographic locations. 
    In this work, our first contribution is to investigate the molecular absorption effect in different THz communication scenarios, including terrestrial, aerial and space, and nanoscale communications. For each scenario, the distribution and the humidity differ and, correspondingly, proper approaches to treating the molecular absorption and designing THz systems are investigated. 

    Apart from the strategies to overcome {the unfavorable consequences brought by} the molecular absorption effect, a remaining problem is whether we can utilize the molecular absorption effect to benefit THz communications. Although the increased attenuation caused by molecular absorption degrades the quality of wireless links, this attenuation is not necessarily harmful if we stand from the perspective of information security. To explore this, our second contribution of this work is to present two novel secure THz communication schemes which utilize molecular absorption to safeguard wireless transmission. Both schemes are motivated by noticing that security aims at not only improving the quality of legitimate link, but degrading or disabling the eavesdropper's link. Therefore, the security of THz communications can be enhanced by designing a  mechanism where the molecular absorption degrades the eavesdropper's link severer than the legitimate one. {By jointly considering the degraded channel capacity and the enhanced security and covertness by the molecular absorption, we claim that it is a double-edged sword of THz communications.}
    
    The remainder of this paper is organized as follows. In Sec.~\ref{sec:mol_abs}, we investigate the molecular absorption effect and its properties. In Sec.~\ref{sec:MA_enlightened}, the molecular absorption enlightened THz communication design for terrestrial, aerial and space, and nano-communications is presented. In Sec.~\ref{sec:MA_secure}, we discuss two security-oriented schemes by utilizing the molecular absorption effect. Sec.~\ref{sec:PE} evaluates the performance of the molecular absorption effect techniques. We elaborate open problems and future research directions in Sec.~\ref{sec:Openproblem}. Finally, the article is concluded in Sec.~\ref{sec:Concl}.

	\section{Molecular Absorption Effect Characterization in the Terahertz Band}
	\label{sec:mol_abs}
\begin{figure*}
	    \centering
	    \includegraphics[width=0.8\textwidth]{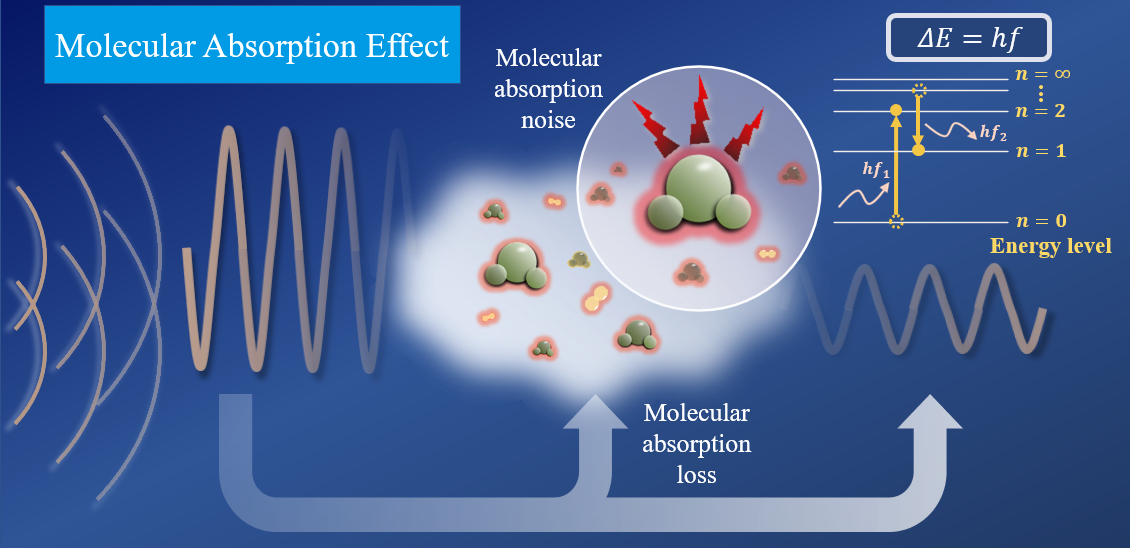}
	    \caption{Demonstration of the molecular absorption effect: Molecular absorption loss and noise.}
	    \label{fig:MA}
	\end{figure*}

	In this section, we first characterize the essence of the molecular absorption effect, which occurs simultaneously with the free-space propagation of an EM wave.
	{Common at any frequency, EM waves emitted from a source spread spherically and experience free-space path loss. During this process, certain types of molecules inside the non-vacuum medium could resonate with, or equivalently, be excited by the THz EM wave. 
	In quantum mechanics, each type of molecule has intrinsic vibrational or rotational energy levels, and the transitions between different energy levels are allowed. A molecule at high energy levels can spontaneously transit into low energy levels by emitting photons, while the energy carried by the photon is equal to the energy level difference. This process also occurs inversely. When external photons with proper energy coincide with molecules, the molecules can transit from low energy to high energy levels, by absorbing the photons. According to the wave-particle duality, the energy of a photon is equal to $hf$, where $h$ denotes the Planck constant, and $f$ represents the frequency of the EM wave. Thus, only EM waves with certain frequencies can be absorbed or emitted by molecules.
	From the perspective of communications, the absorbance of a photon leads to the attenuation of the EM wave, and the emission of photons brings an additional noise in THz channels. This phenomenon is referred to the molecular absorption effect, as illustrated in Fig.~\ref{fig:MA}.}
    
    Beyond the reference model for the molecular absorption loss between 0.1-1 THz range given by the ITU recommendation P.676-12~\cite{ITUR}, in~\cite{jornet2011channel}, a comprehensive model for both the molecular absorption loss and the molecular absorption noise for the entire 0.1-10~THz range was provided. Generally, molecular absorption loss is modeled by the summation of the power absorbed by all of the molecules along the propagation path. Notably, the molecular absorption loss can be calculated as the product of the molecular absorption coefficient $k$ in $\textrm{dB}/\textrm{km}$ and the propagation distance $d$ in $\textrm{km}$. In the THz band, water vapor leads to the most significant molecule absorption effect among all types of common air components. Quantitatively, \textit{water vapor} dominates the molecular absorption, with six orders of magnitude larger than the second contributor, \textit{oxygen}~\cite{jornet2011channel}.
    Moreover, the coefficient $k$ has \textit{triple selectivity}, i.e., it exhibits a strong dependence on the frequency, the distance, and various environmental parameters including pressure, temperature, and medium composition.
    Due to such triple selectivity, it is critical to study the features of the molecular absorption effect and use them to enlighten the design of THz communication systems, as follows.

\begin{figure*}
	    \centering
	    \includegraphics[width=0.8\textwidth]{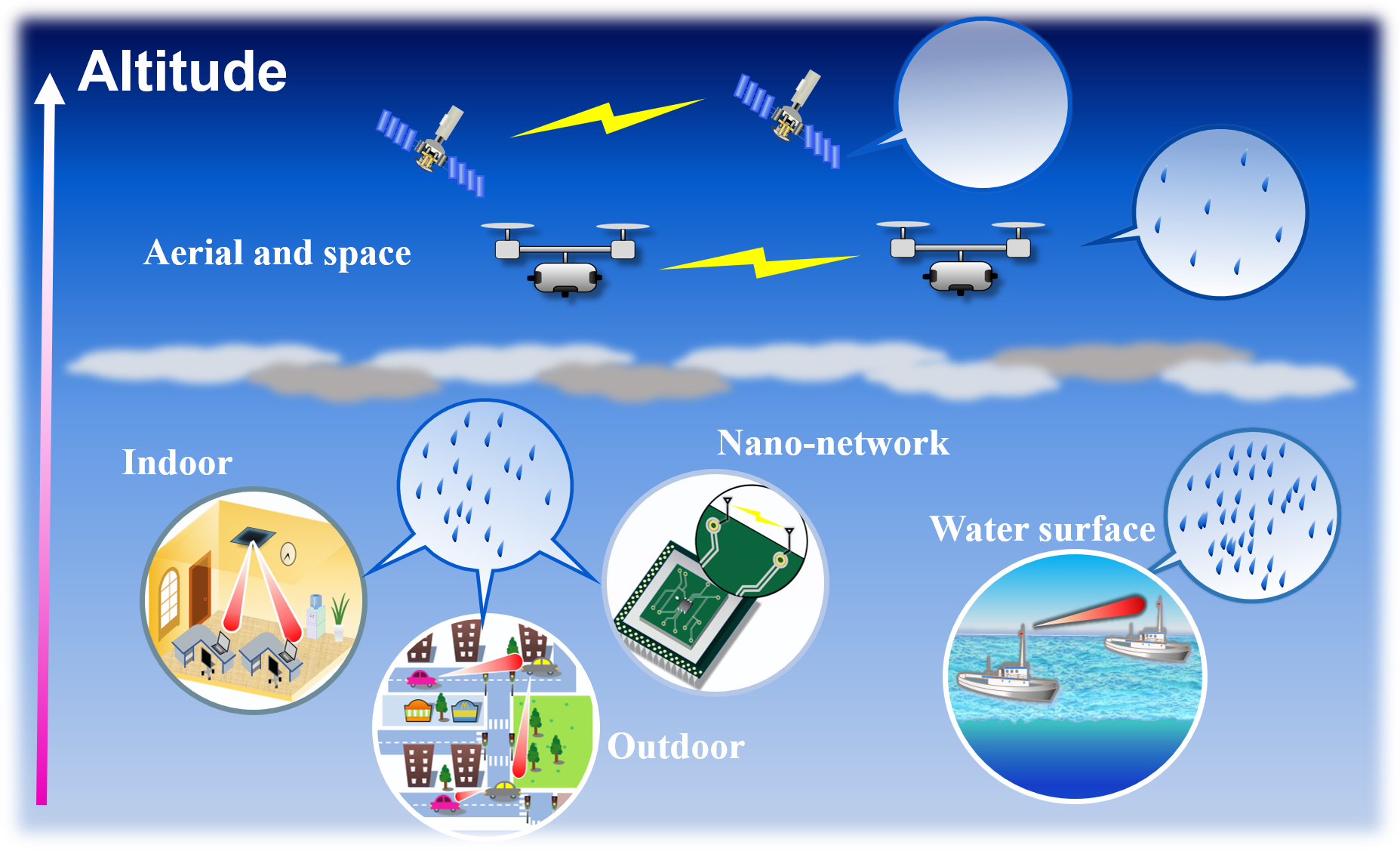}
	    \caption{Terahertz wireless communication scenarios. }
	    \label{fig:scene}
	\end{figure*}

    \begin{itemize}
        \item \textbf{Frequency selectivity:} The molecular absorption effect depends on the intensity of the resonance between EM waves of certain frequencies and medium molecules, which makes the molecular absorption highly frequency-selective. At resonant frequencies, the molecular absorption effect is locally maximized and many molecular absorption peaks are created in the THz band, between which the THz spectral windows that own multi-GHz bandwidths are defined. Wireless transmission performed within these spectral windows is suggested to avoid experiencing severe molecular absorption loss. To this end, multi-wideband transmission occupying non-contiguous spectral windows was proposed in~\cite{7321055}.  
        \item \textbf{Distance dependence:} In a homogeneous medium, the molecular absorption loss exponentially increases with the propagation distance. {For an inhomogeneous medium, the exponential increase in molecular absorption loss is the same, but the speed of this increase changes with the medium components.} In both mediums, when the propagation distance increases, the molecular absorption loss firstly appears negligibly weak, but then increases explosively and rapidly with the distance. This phenomenon would strictly limit the coverage of THz communications if the transmission scheme does not adapt to the distance. As a countermeasure, distance-adaptive communication was suggested in~\cite{hossain2019hierarchical}, which adjusts the transmission frequency band dynamically with the distance to avoid distance-dependent absorption peaks. 
        \item \textbf{Environment impact:}
        Since the molecular absorption effect is closely related to the water vapor concentration along the propagation path, its impact on THz communications significantly varies under different communication environments. Typical THz communication scenarios are summarized in Fig.~\ref{fig:scene}, including terrestrial, aerial and space, and nano-scale communications. {From low altitude to high altitude regions, the concentration of water vapor decreases, which significantly affects the design guidelines related to molecular absorption.}
        Specific communication design guidelines to address different molecular absorption properties are elaborated in the following section.
\end{itemize}        
    Apart from attenuation, molecular absorption further introduces noise, in addition to the common thermal noise in wireless systems. The modeling of the molecular absorption noise was described in~\cite{jornet2011channel,jornet2014femtosecond}. The main feature of the molecular absorption noise is its dependency on the transmitted signal. Specifically, at certain frequency, distance, and environment, the molecular absorption noise is linearly dependent on the transmit power of the signal. This feature motivates the asymmetric bit-symbol mapping, especially for pulse-based modulation in nano-scale communications~\cite{jornet2014low}, as elaborated in Sec.~\ref{sec:nano}.

	\section{Molecular Absorption Enlightened Terahertz Communications}\label{sec:MA_enlightened}
\subsection{Terrestrial Terahertz Communications}
	Indoor and outdoor are two categories of terrestrial communication scenarios. 
	For indoor terrestrial THz communications, the stability of environmental parameters brings great merits to communications. First, an indoor scenario is usually immune from the influence of weather, and the concentration of water vapor in indoor environments changes insignificantly on rainy or snowy days. Second, the indoor pressure, temperature, and moisture keep in a relatively stable condition. Third, the maximum propagation distance of indoor wireless links is confined by the size of the indoor environment. 
	Owing to these indoor characteristics, the indoor molecular absorption coefficient is upper bounded, and thus an indoor THz system design does not need to consider the distance adaptiveness and environment dependency of the molecular absorption effect. 
	Moreover, the indoor properties suggest that it may be unnecessary to consider the location dependency of the molecular absorption effect, since the difference among various geographical locations is inconsiderable under indoor environments. 
	
	For outdoor terrestrial communications, however, weather impact becomes a critical challenge when designing THz communication systems~\cite{federici2016review}. Unlike indoor communications, the density of water vapor in outdoor environments significantly changes under certain weather conditions such as rain and snow. For example, in heavy rain, the concentration of water vapor dramatically increases, making more THz spectral bands unavailable to communicate. This brings a design constraint that on rainy days, THz communications undergo severe attenuation and the link quality is very likely to be severely degraded.

	\subsection{Aerial and Space Terahertz Communications}
	In aerial and space communication scenarios such as low altitude platforms and high altitude platforms, the molecular absorption effect caused by the water vapor and oxygen in the air is highly dependent on altitude and weather.
    On one hand, the major factor affecting THz aerial and space communication systems is altitude.  The pressure and temperature of the air, as well as the density of water vapor, vary significantly at different altitudes. A critical borderline occurs between the stratosphere and the troposphere, at around 10~km above the sea level. In the troposphere regions with altitudes less than 10~km, the water vapor significantly decreases as altitude increases, and the molecular absorption rapidly decreases. By contrast, in the stratospheric region with altitude above 10~km, water vapor almost disappears and oxygen dominates the molecular absorption. When altitude further increases from the stratosphere region to the vacuum space, there is no molecular absorption effect anymore, and the molecular absorption effect is much less significant than that in terrestrial and troposphere regions~\cite{kokkoniemi2021channel}.
	
\begin{figure}
    	\centering
    	\includegraphics[width=\textwidth/2]{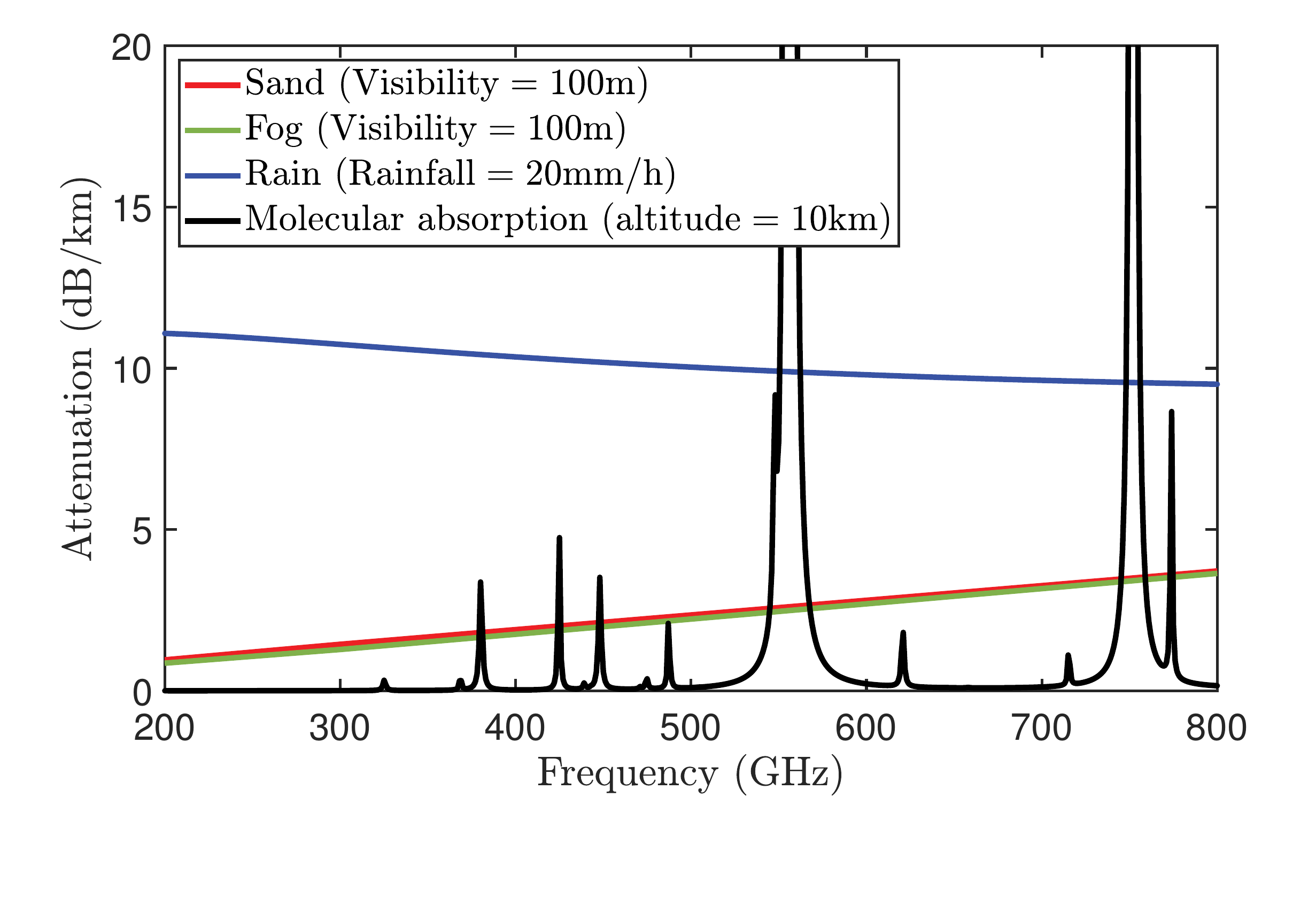}
    	\caption{Attenuation versus frequency for the different weather conditions.}
    	\label{fig:weather}
    \end{figure}

	On the other hand,  weather is another key factor affecting aerial THz communications. On rainy or snowy days, the concentration of water vapor increases, which enlarges the influence of the molecular absorption effect. We should notice that since these weather conditions occur in the troposphere region, the weather only affects low altitude aerial THz communications. 
	Furthermore, weather not only degrades the transmission by increasing the molecular absorption effect, but causes scattering loss. In sand and dust weather, the solid particles mixed in the air scatter the encountered EM waves and result in additional propagation loss. {Fig.~\ref{fig:weather} shows the path loss caused by various weather conditions and the molecular absorption in the THz band. We observe that the molecular absorption demonstrates a significant frequency selectivity. Moreover, rain contributes to THz wave attenuation more significantly, in contrast with sand and fog.}
    \subsection{Terahertz Sea Surface Communications}
    Sea surface communications play an important role in inter-ship and low-airborne-altitude communication scenarios over the sea. Due to the condition that seawater steadily evaporates, water vapor is almost saturated along the over-the-sea links all the time. Moreover, the water vapor saturation condition is almost permanent for sea surface communication, which is different from terrestrial outdoor communication scenarios where it only occurs on rainy days. To address this disadvantage, visible light communications (VLCs) are envisioned as a good compensation for THz communications, as the link quality of VLCs is not severely degraded by water vapor~\cite{tabassum2018coverage}.
	
	\subsection{Nano-scale Terahertz Communications}\label{sec:nano}
	Nanonetworks have attracted great attention to realize nano-scale applications such as health-monitoring systems and the Internet of Nano-Things (IoNT). New nano-antennas are found to be more efficient in emitting pulse signals than carrier signals, which  occupies a contiguous large piece of spectral band~\cite{jornet2013graphene}.
    Thus, the transmitted signal is inevitable to cover frequency bands that contain molecular absorption peaks. However, due to the extremely short distance, the majority of the THz transmission power does not experience significant molecular absorption attenuation. 
	
	Instead, it is necessary to consider the molecular absorption noise for pulse-based nano-scale modulation schemes. According to the fact that molecular absorption noise linearly depends on the transmit power of the main signal, an asymmetric bit-symbol mapping design is motivated. 
	{Transmitting a pulse signal and a silent signal results in different received signal-to-noise ratios (SNRs), since transmitting a pulse signal excites an additional molecular absorption noise. Therefore, a wise strategy of symbol design is to transmit more silent signals than pulse signals. To this end, the time spread on-off keying (TS-OOK) modulation and low-weight error-prevention coding schemes for THz nanonetworks were proposed in~\cite{jornet2014femtosecond,jornet2014low}, which demonstrate that transmitting approximately 45\% ones and 55\% zeros reaches the maximum capacity. }
	
	\section{Molecular Absorption Aided Secure and Covert Communications}\label{sec:MA_secure}
	From the perspective of increasing data rates, the attenuation and noise caused by the molecular absorption effect are harmful. For this reason, most existing THz communication systems avoid using frequency bands around molecular absorption peaks. Despite so, an interesting problem is whether it is possible to take the advantage of molecular absorption effect to benefit communications. We endeavor to offer a positive answer: \textit{Molecular absorption aided secure and covert communications}. 
	
	{We clarify that when any of the secure and covert schemes proposed in this section is applied to one of the scenarios mentioned in Sec. III, such as terrestrial, aerial and space, sea surface, or nano-scale scearios, the frequency-dependent molecular absorption effect is first formulated for each scenario. Then, the carrier frequency of the system is optimized for each scenario by jointly considering the legitimate link quality issue and the security and covertness requirement.}

    \subsection{Challenges of Terahertz Secure Communications}
	Information security is one of the critical demands in today’s communication systems and the trend grows.
	Due to the drawbacks of traditional cryptography technologies such as large overhead, new security mechanisms exploiting physical layer characteristics is awaited. Moreover, covertness is gaining growing attention as a more stringent requirement than conventional security since it protects the existence of signal transmissions itself, rather than the carried information. 
	
	Different from the conventional communication requirement of achieving high data rates, security and covertness targets to not only ensure the SNR at the legitimate user (LU) meeting reliability requirement but reduce the SNR of the signal detected by eavesdroppers below a certain threshold. Particularly, for covertness,  the transmitter should guarantee an even lower SNR at the eavesdropper to conceal the transmission under the background noise.
	These stringent requirements of maintaining a high-quality legitimate channel and suppressing a low eavesdropper capacity lead to the key security design methodology, namely, widening the capacity gap between the LU and the eavesdropper. 
	
    Aligning with the key security design methodology, the THz spectrum is envisioned to possess enticing potential to support security and covertness.
    Since THz transmissions experience a high spreading loss according to the Friis' law, a high-gain and narrow beam is usually created by ultra-massive multiple-input multiple-output (MIMO) to realize directional transmission towards the LU. 
    Thus, an eavesdropper residing outside the narrowbeam is hard to overhear the transmission, while the LU is benefited by the antenna gain~\cite{ma2018security}. 
    However, this directivity-aided mechanism is not always reliable, especially when an eavesdropper moves inside the narrowbeam. Unfortunately, it is impractical to acquire the positions of all eavesdroppers, which makes the directivity-aided mechanism ineffective~\cite{myarticle1}.

    \subsection{Molecular Absorption Peak Modulation}
    Motivated by its exponential distance-dependent feature, introducing molecular absorption loss into secure communications can slightly degrade the legitimate link while greatly mitigating the eavesdropper's link capacity, when the LU is nearer from the transmitter than the eavesdropper. 
    Based on this strategy, molecular absorption peak modulation (APM) can be designed by considering the following two constraints.
    First, the molecular absorption loss at the LU is limited, to support reliable transmission. Second, for security and covertness, the molecular absorption loss for eavesdroppers should be as large as possible, such that the received SNRs at eavesdroppers are significantly degraded. 
    {Thus, there is a trade-off between introducing a large molecular absorption loss for the sake of security and a small molecular absorption loss for ensuring the reliability of the legitimate link~\cite{myarticle1}. 
    When selecting the frequency bands around molecular absorption peaks, the carrier frequency of the system, which determines the molecular absorption loss, is optimized for the scenario where THz communication is applied, by jointly considering the legitimate link capacity and the security requirement.}
    
    For the drawback of the APM scheme, due to the exponential function of the molecular absorption loss it can only work effectively when the eavesdropper is located farther than the receiver. Moreover, the data rate of the LU is partially sacrificed by introducing an additional molecular absorption loss. Therefore, this scheme cannot be applied when the legitimate link itself is weak and unstable. 
    
    \subsection{Molecular Absorption Aided Receiver Artificial Noise Transmission}
    In the case of eavesdroppers in close proximity where  eavesdroppers are nearer from the transmitter than the LU, 
    a widely-used idea is to simultaneously transmit artificial noise (AN) signals, which do not bear information, together with information signals to confuse the eavesdroppers. Most existing studies have suggested the transmitter to generate AN signals, which nevertheless are not applicable for THz line-of-sight (LoS) transmission, since the LoS path is shared between the LU and the eavesdroppers. An alternative method is to allow the full-duplex receiver to transmit AN signals while receiving information, which is known as receiver AN (RAN).
    In the traditional implementation of RAN, the self-interference (SI) caused by AN signal at the LU is not negligible. To address this problem, the receiver is required to cancel the SI by deploying some SI cancellation mechanisms, which brings high hardware difficulty and is impractical for the state-of-the-art THz technologies.
    
    Against this background, a novel molecular absorption aided RAN scheme was proposed in the THz band~\cite{myarticle2}, where SIC-free transmission is realized through an interesting waveform design based on the temporal broadening effect (TBE) caused by molecular absorption.  The TBE refers to the widening of transmitted pulses that experiences frequency-selective attenuation~\cite{myarticle2}. An important phenomenon is observed that when pulses are transmitted in frequency bands with large molecular absorption, the TBE increases considerably with distance. 
    In light of this observation,  the SI can be easily removed without incurring any complicated compensation-based SIC methods. Specifically, at the LU, since the SI signal caused by transmitting AN signals is very close to its receive antennas, the SI experiences negligible TBE. Motivated by this, we can separate information signals and AN signals in the time domain while keeping these signals overlapping at eavesdroppers after experiencing TBE, since the duration of received SI is extremely short at the LU. As a result, the LU can detect information signals without any influence from AN signals. Meanwhile, the information signals at the eavesdropper are heavily polluted by the AN signals. Therefore, accounting for proximal eavesdroppers, the proposed RAN scheme can guarantee THz secure communications without the need of SIC.

	\section{Performance Evaluation}\label{sec:PE}
\begin{figure}
	    \centering
	    \includegraphics[width=\textwidth/2]{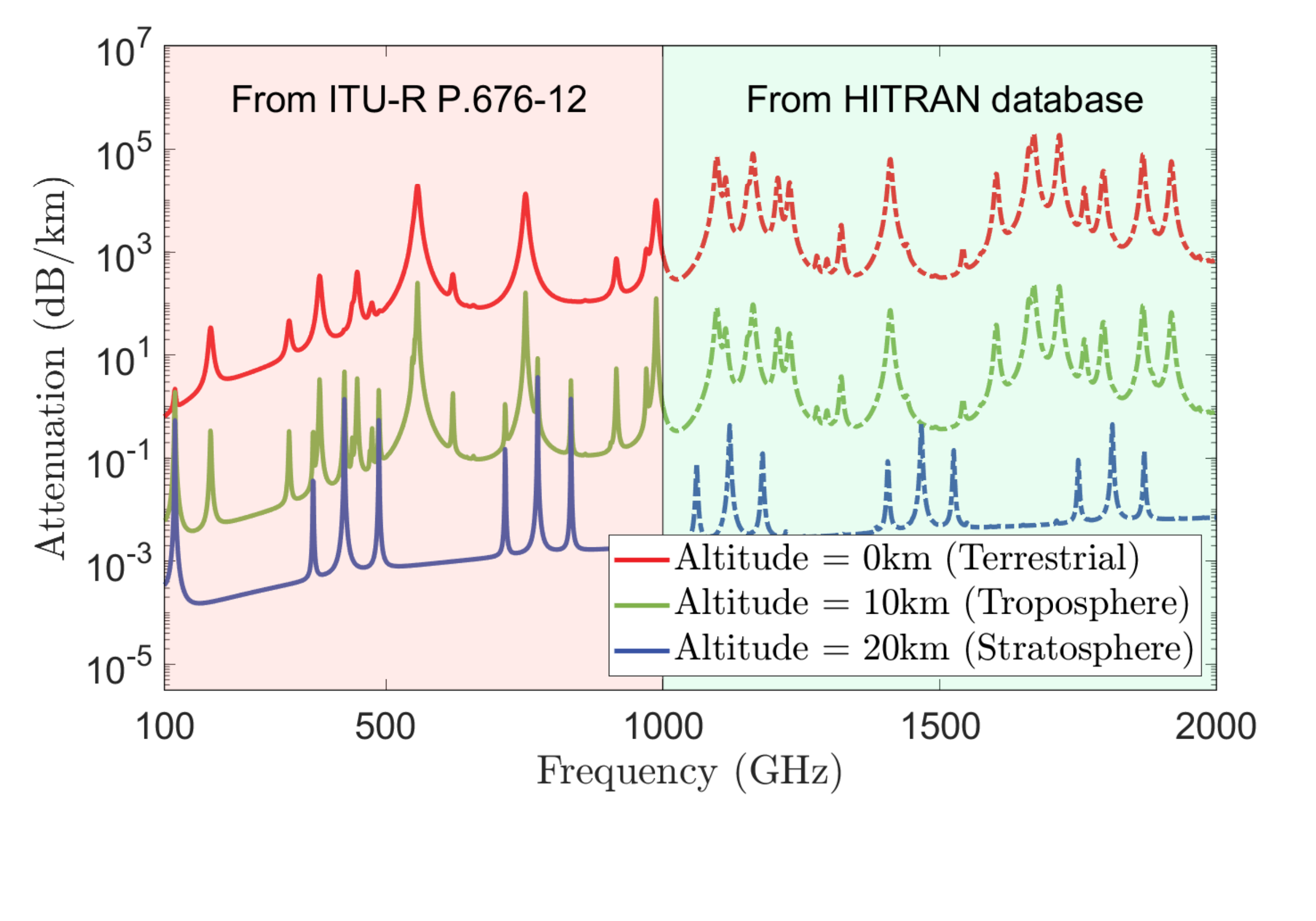}
	    \caption{The molecular absorption loss versus frequency with altitude at 0~km, 10~km, and 20~km, corresponding to the terrestrial communication, troposphere and stratosphere in summer and high latitude region. }
	    \label{fig:MA_alt}
	\end{figure}

	We numerically evaluate the molecular absorption loss to illustrate its significance in designing THz communications. In Fig.~\ref{fig:MA_alt}, the molecular absorption attenuation in the frequency range from 100~GHz to 2~THz is plotted for different altitudes. The simulation parameters are based on the standard pressure, temperature, and water vapor concentration data in summer and high latitude areas, as $1~\textrm{atm}$, $290~\textrm{K}$, and $7.5~\textrm{g}/\textrm{m}^3$, respectively. The molecular absorption loss values from $100~\textrm{GHz}$ to $1~\textrm{THz}$ are based on the data in ITU-R P.676-12, while the values from $1~\textrm{THz}$ to $2~\textrm{THz}$ are calculated based on the data in the high-resolution transmission molecular absorption database (HITRAN)~\cite{jornet2011channel}. We discover that the molecular absorption decreases at approximately two orders of magnitude per kilometer as the altitude increases. For terrestrial and aerial communications in the troposphere, the molecular absorption loss shows the same shape, which indicates that the water vapor dominates the molecular absorption. However, for aerial communications in the stratosphere, oxygen becomes the main factor affecting the molecular absorption effect.
	
\begin{figure}
	    \centering
	    \includegraphics[width=\textwidth/2]{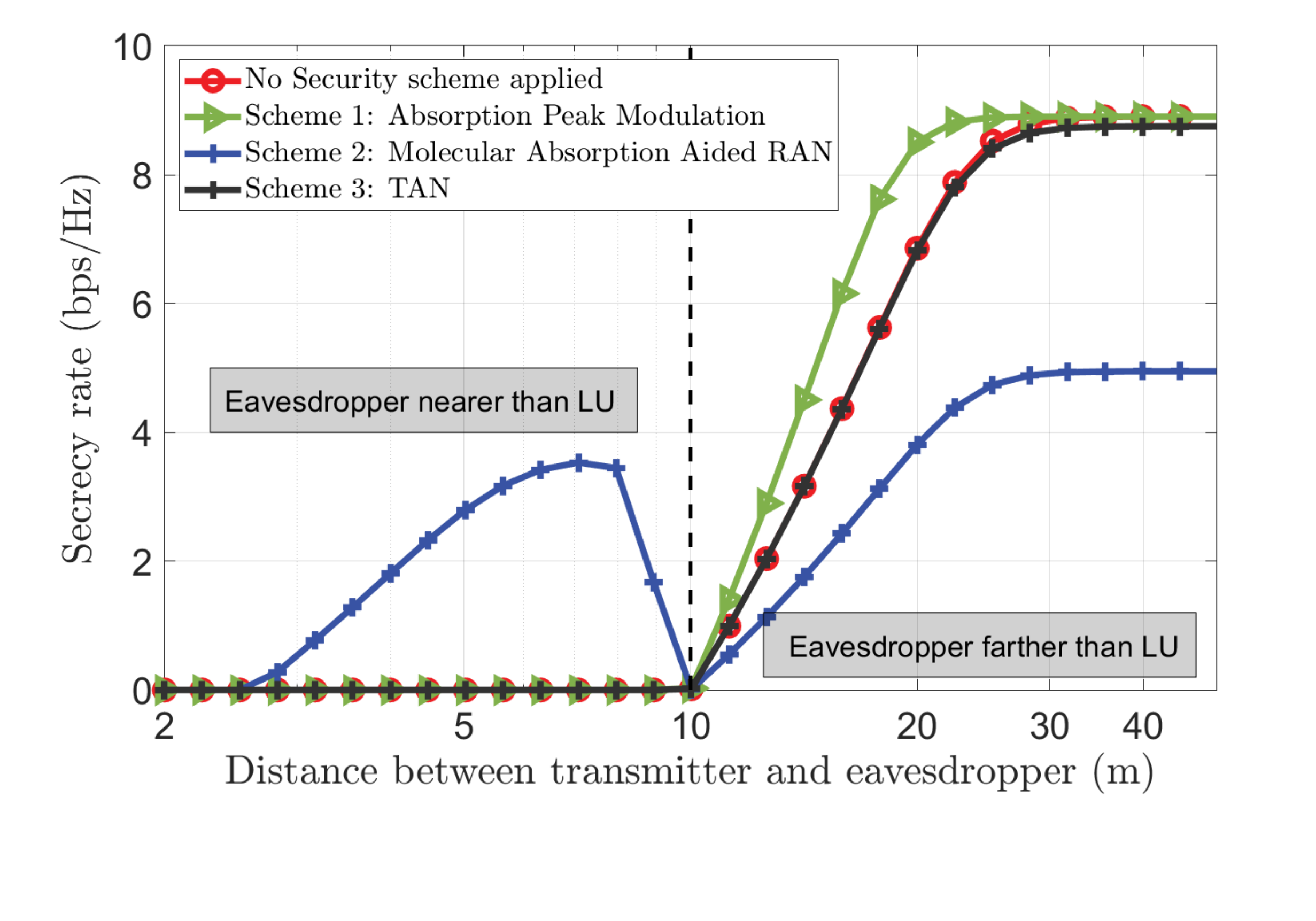}
	    \caption{{Quantitative performance comparison of the proposed security techniques.}}
	    \label{fig:Rs}
	\end{figure}

    Next, we demonstrate the security performance enhancement brought by two molecular absorption enlightened secure communications schemes in Fig.~\ref{fig:Rs}. The secrecy rate, defined as the maximal data rate which can be transmitted reliably and confidentially, is adopted as the security measure. In our simulation, we fix the LU to be located 10~m away from the transmitter and vary the transmitter-eavesdropper distance from 2~m to 50~m. As a benchmark, the red curve represents the secrecy rate without adopting any security measures. {The black curve stands for the secrecy rate achieved by the transmit AN scheme, where the transmitter side simultaneously transmits the AN signal and the legitimate signal. It is shown in the figure that the TAN scheme cannot combat eavesdroppers nearer than LU.}
    When the transmitter-eavesdropper distance is smaller than 10m, the SIC-free scheme can achieve the maximum secrecy rate of $3.8~\textrm{bps}/\textrm{Hz}$, becoming the sole scheme to achieve a positive secrecy rate. 
	{When the transmitter-eavesdropper distance equals $10~\textrm{m}$, the location of receiver overlaps with that of the eavesdropper. As a result, the information transmitted to the receiver is inevitably decodable or detectable by the eavesdropper, which leads that the secrecy rate reduces to zero.}
    
    Furthermore, the drawback of the APM and SIC-free schemes is also reflected. First, the APM scheme does not work when the eavesdropper is in close proximity of the transmitter~\cite{myarticle1}. Second, the SIC-free RAN scheme may degrade the security when the eavesdropper is farther than the LU~\cite{myarticle2}.    
    \section{Open Problems And Potential Research Directions}\label{sec:Openproblem}
    \subsection{Modeling of Spectral Windows}
     {Curve fitting methods have been commonly used to approximate the frequency-dependent molecular absorption coefficient. However, most studies have focused on the fitting performance of the peaks of original curves. For example, the model in~\cite{jornet2011channel} accurately captures molecular absorption peaks, while fitting the rest of the curve by deploying an asymmetric line shape model. Nevertheless, the existing approximation method may significantly under- or over-estimate the ``flat'' regions of the molecular absorption. Since THz signals are transmitted in ``flat'' regions that emerge the spectral windows, a precise molecular absorption fitting model for the entire spectral windows is required. 
     Moreover, we have demonstrated that spectral windows highly depend on altitude. Under different communication scenarios, especially at various altitudes, the water vapor distribution, as well as the molecular absorption effect, differs a lot. Due to this, the characterization of spectral windows should be altitude-dependent to make the maximal use of the THz spectrum resource.}

    \subsection{{Hybrid Visible Light and Terahertz Communications}}
    Motivated by their different transmission features, hybrid visible light communication and radio frequency (VLC/RF) systems have been proposed to cooperate and overcome the challenges of two individual communication systems~\cite{tabassum2018coverage}. For example, VLC communications heavily rely on the LoS path whereas being robust against multipath fading, while RF transmissions suffer greatly from multi-path fading and the necessity of the LoS path is not strict. Motivated by such facts, hybrid VLC and THz communications can be explored as a promising direction, for the following reasons. Owing to the fact that water vapor is the major cause of THz wave attenuation, THz transmission is unreliable during rain and snow with high humidity. On the upside, THz communications are barely affected by air particles resulting from clouds, sand, and dust.
	By contrast, VLC links are immune from water vapor absorption, whereas being vulnerable to scattering by air particles. 
	Based on these characteristics, hybrid VLC/THz communication systems can take advantages of their communication abilities in various weather conditions to improve the overall transmission quality.
	\subsection{Security for Proximal Receiver and Eavesdropper}
\begin{figure}
	    \centering
	    \includegraphics[width=\textwidth/2]{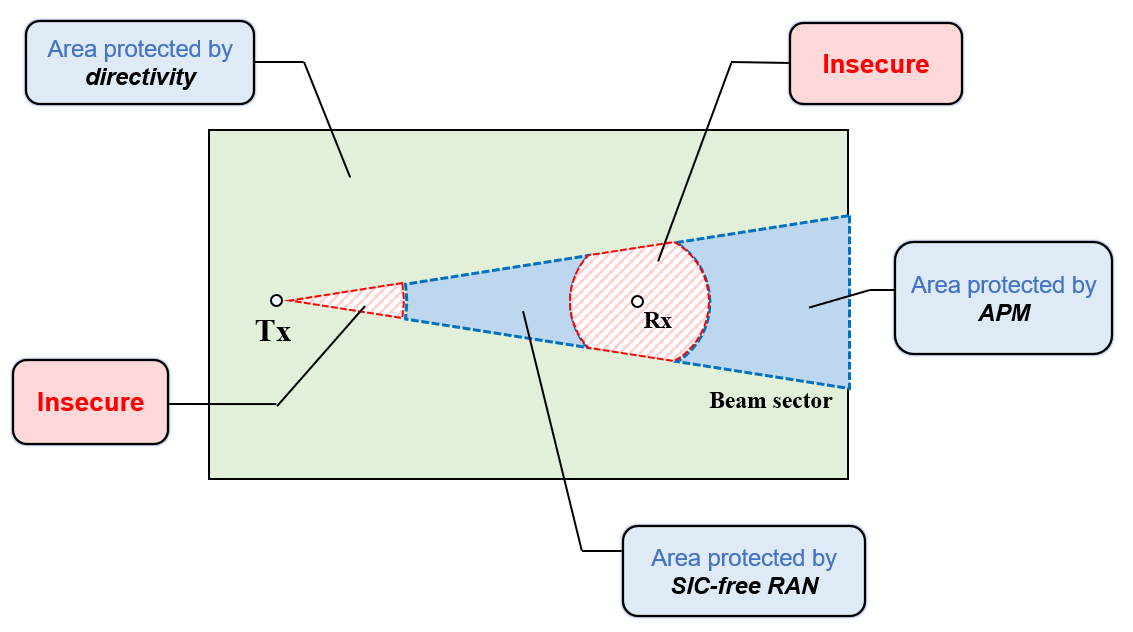}
	    \caption{Demonstration of security strategies applied for eavesdroppers in different locations.}
	    \label{fig:sec}
	\end{figure}

	Among all possible eavesdropper locations, we have demonstrated that i) the outside-beam area, ii) the far and inside-beam area, and iii) near and inside-beam area can be protected by the directivity and the two molecular absorption enlightened secure communication schemes, respectively, as shown in Fig.~\ref{fig:sec}. 
	{There remain two insecure regions. First, there is an insecure zone around transmitter, where the path gain of the transmitter-eavesdropper link is very large according to Friis’ law. In this region, it is difficult to consider physical layer security tools. Second, the insecure region zone around receiver is due to the high channel correlation between the transmitter-receiver and the transmitter-eavesdropper links. When the two channels are highly correlated, the information transmitted to the receiver is inevitably decodable or detectable by the eavesdropper. Hence, } The angle-dependent directivity and range-dependent molecular absorption effect cannot work effectively.
	
	Some recent studies have used a frequency diverse array (FDA) to solve this proximal receiver and eavesdropper problem~\cite{lin2017physical}, where the FDA generates a range-varying beam pattern to create a capacity difference between the two proximal locations.
	However, the illustration of the beam pattern shows a ``snapshot" spatial distribution of the beamforming gain, and thus the capacity difference is incorrectly calculated, by neglecting the distance between the receiver and the eavesdropper. Unfortunately, this capacity difference is not the correct secrecy rate, since the same piece of signal does not reach the LU and eavesdropper at different distances simultaneously. 
	Instead, widely-spaced antenna communications are motivated to combat this problem, thanks to its range-dependent array pattern created by spherical-wave propagation nature. Further analysis of this method and relevant security-enhancing strategies are awaited to be explored.
	\section{Conclusion}\label{sec:Concl}
    In this article, we focused on the molecular absorption effect, which is a key THz spectrum feature that makes the THz band distinctive from other frequency bands.
    Specifically, how the molecular absorption effect affects THz communication system design under various scenarios, including terrestrial, aerial and space, and nano-communications, has been analyzed and numerically evaluated. 
    As the other side of the double-edged sword, molecular absorption could be beneficial in THz secure and covert communications. Motivated by this, a molecular absorption peak transmission scheme and a molecular absorption aided SIC-free RAN scheme have been presented and their benefits for security have been examined.
		{The modeling of the altitude-dependent molecular absorption effect and the spectral windows remains an open problem. Moreover, new opportunity of utilizing the molecular absorption to enable hybrid VLC/THz communications is awaited to be explored.}
    %Finally, open problems and future directions for molecular absorption enlightened THz communications design have been identified and discussed.  
	\bibliographystyle{IEEEtran}
	\bibliography{main_rev}
\end{document}